\documentclass[12pt]{article}
\def\newpic#1{}

\input{epsf}

\newtheorem{theorem}{Theorem}

\newenvironment{proof}{{\bf Proof.}}{\hfill\rule{2mm}{2mm}}

\newtheorem{remarka}{Remark}

\def\Unique {{\exists ! } }
\def\false {{\mathit false }}
\def\true {{\mathit true }}
\def\SAT {{\rm SAT}}
\def\NP {{\bf NP}}
\def\P {{\bf P}}
\def\supp {{\rm supp}}
\def\CoNP {{\bf CoNP}}
\def\vc { {\sc Vertex coloring} }
\def\sg {${\bf \Sigma_2}$}
\def\bx {{\bf x}}
\def\bv {{\bf v}}
\def\by {{\bf y}}

\def\bw {{\bf w}}

\title{\bf On the Computational Complexity of Defining Sets\footnote{This research
is done in Discrete Mathematics
Laboratory of Department of Mathematics in Sharif University of
Technology.}}

\author{Hamed Hatami\footnote{
Department of Computer Science,
University of Toronto,
 email: hamed@cs.toronto.edu}
\and
Hossein Maserrat\footnote{
Department of Mathematical Sciences,
Sharif University of Technology,
Tehran, Iran}
}

\date{}
\begin{document}
\maketitle
\begin{abstract}
Suppose we have a family ${\cal F}$ of sets. For every $S \in
{\cal F}$, a set $D \subseteq S$ is a {\sf defining set} for
$({\cal F},S)$ if $S$ is the only element of $\cal{F}$ that
contains $D$ as a subset. This concept has been studied in
numerous cases, such as vertex colorings, perfect matchings,
dominating sets, block designs, geodetics, orientations, and Latin
squares.

In this paper, first, we propose the concept of a defining set of
a logical formula, and we prove that the computational complexity
of such a problem is \sg-complete.

We also show that the computational complexity of the following
problem about the defining set of vertex colorings of graphs is
\sg-complete:
\begin{itemize}
\item[]
{\sc Instance:} A graph $G$ with a vertex coloring $c$ and an
integer $k$.
\\
{\sc Question:} If ${\cal C}(G)$ be the set of all
$\chi(G)$-colorings of $G$, then does $({\cal C}(G),c)$ have a
defining set of size at most $k$?
\end{itemize}

Moreover, we study the computational complexity of some other
variants of this problem.
\end{abstract}
{{\sc Keywords:} defining sets; complexity; graph coloring;
satisfiability.

\section{Introduction}

In this paper we consider a unification of the concepts already
known as critical sets, forcing sets, and defining sets, where we
formulate different natural problems in this regard. Specially,
through considering such problems for 3\SAT, by introducing
suitable reductions, we prove that the decision problem related to
the minimum defining set problem of graph
colorings\footnote{Defined formally below} is \sg-complete.

Defining sets were studied for Latin
squares~\cite{MR2000g:05034,MR85k:68035}, perfect
matchings~\cite{AdamsMahdianMahmoodian,AfshaniHatamiMahmoodian,MR94b:05193},
orientations~\cite{MR96m:05094}, geodetics~\cite{MR2000d:05037},
vertex colorings~\cite{MR98b:05044}, designs~\cite{MR91e:05016},
and dominating sets~\cite{MR98f:05083}.

Let ${\cal F}$ be a family of sets. For every $S \in {\cal F}$, a
set $D \subseteq S$ is a {\sf defining set} of $({\cal F},S)$, if
$S$ is the only element in ${\cal F}$ which contains $D$ as a
subset. By abuse of language every defining set of $({\cal F},S)$
is also called a defining set of ${\cal F}$.

In what follows, we try to introduce a general formulation for
the type of problems we are going to consider in the rest of this
paper.

Suppose an input $I$ is given. The input $I$ might be a graph, a
number, or any other mathematical object. Then let ${\cal F}(I)$
be a family of sets which is defined according to the set $I$. In
this paper we are interested in the computational complexity of
the following three general types of questions for specified
inputs and definitions of ${\cal F}$.

\begin{enumerate}
\item
\begin{itemize}
\item Q1
\item[] {\sc Instance:} $I$, a set $S \in {\cal F}(I)$, and a set $D \subseteq S$.
\\
{\sc Question:} Is $D$ a defining set of $({\cal F}(I),S)$?
\end{itemize}
\item
\begin{itemize}
\item Q2 \item[] {\sc Instance:} $I$, a set $S \in {\cal F}(I)$,
and an integer $k$.
\\
{\sc Question:} Does $S$ have a defining set of size at most $k$?
\end{itemize}
\item
\begin{itemize}
\item Q3
\item[] {\sc Instance:} $I$ and an integer $k$.
\\
{\sc Question:} Does ${\cal F}(I)$ have a defining set of size at
most $k$?
\end{itemize}
\end{enumerate}

The computational complexity of the problems related to defining
sets was first studied by Colbourn in~\cite{MR85k:68035}. He
studied Q1 when ${\cal F}(n)$ is the set of Latin squares of order
$n$, and proved that this question is \CoNP-complete. Recently
Adams, Mahdian, and Mahmoodian~\cite{AdamsMahdianMahmoodian}
studied Q2 when ${\cal F}(G)$ is the set of perfect matchings of a
graph $G$, and proved that the question is \NP-complete.
In~\cite{AfshaniHatamiMahmoodian} it is shown that the question Q3
for this family is \NP-complete. It is not hard to see that the
question Q1 for this family is in \P. Hatami and Tusserkani
in~\cite{HatamiTusserkani} studied Q2 and Q3 when ${\cal F}(G)$ is
the set of vertex colorings of a graph $G$, and proved that both
of the questions are \NP-hard. In this paper we improve their
result by showing that these problems are both \sg-complete. In
this regard we consider the family of all proper assignments
 to the variables of a $k$CNF where a $k$CNF is a Boolean
expression in conjunctive normal form such that every clause has
exactly $k$ variables. Let Q1-$k\SAT$, Q2-$k\SAT$, and Q3-$k\SAT$
stand for the three questions Q1, Q2, and Q3 in this case,
respectively. We show that Q1-$3\SAT$ is \CoNP-Complete, and
Q2-$3\SAT$ and Q3-$3\SAT$ are both \sg-complete. We also refer the
reader to the recent paper~\cite{Oleg} for some other
computational complexity results on the defining sets of vertex
colorings.

We determine the computational complexity of Q1-$3\SAT$,
Q2-$3\SAT$, and Q3-$3\SAT$ in Section~2. Section~3 is devoted to
the study of the computational complexity of the questions Q1, Q2,
and Q3 for the set of vertex colorings of a graph.

\section{Defining sets and SAT}
Let $D$ and $R$ be two sets, and $f:D\rightarrow R$ be a function.
We can refer to $f$ as the set $f=\{(x,f(x)):x \in D\}$. This
representation enables us to study the defining set of a family of
functions.

Let $\phi$ be a $k$CNF with variables $V=\{ v_1, v_2, \ldots,
v_n\}$. For the sake of simplicity we use the notation $\phi(\bv)$
instead of $\phi(v_1,\ldots,v_n)$, where we would think of $\bv$
as a vector of $v_1, v_2, \ldots, v_n$. Since any truth assignment
$t:V \rightarrow \{\true,\false\}$ is a function, we can study the
defining set of a family of assignments. A {\sf proper assignment}
of $\phi$ is an assignment which makes $\phi$ true. Let $S
\subseteq V$, be a subset of the variables of $\phi$. A {\sf
partial assignment} of $\phi$ over the set $S$ is a truth
assignment $t:S \rightarrow \{\true,\false\}$. The set $S$ is
called the support set of $t$, and this is denoted by
$S=\supp(t)$. A partial assignment over $S$ is called {\sf proper}
if every clause of $\phi$ contains at least one true literal from
the variables in $S$.

For every $k$CNF $\phi$, let ${\cal P}(\phi)$ denote the family of
proper assignments of $\phi$. We study the computational
complexity of the general questions Q1, Q2, and Q3 for this
special family. Let Q1-$k\SAT$, Q2-$k\SAT$, and Q3-$k\SAT$ stand
for the three questions Q1, Q2, and Q3 in this case, respectively.

Duplicating a variable in a clause of a CNF  does not change the
family of its proper assignments. Hence if all clauses of a CNF
are of size at most $k$ (not necessarily equal to), then it can be
converted to a $k$CNF. Therefore without loss of generality, we
may always assume that all such expressions are in $k$CNF form.

In this section we show that Q1-$3\SAT$ is \CoNP-Complete, and
Q2-$3\SAT$ and Q3-$3\SAT$ are both \sg-complete.
From~\cite{Wrathall} we know that the following problem is
\sg-complete.

\begin{itemize}
\item $\exists {\not\exists} \ 3\SAT$ \item[] {\sc Instance:} A
3CNF, $\phi(\bx,\by)$.
\\
{\sc Question:} Is $\exists \bx {\not\exists} \by \
\phi(\bx,\by)$?
\end{itemize}
Next, we define the $\exists \Unique_* \ k\SAT$ problem, and prove
that it is \sg-complete. A $k$CNF, $\phi$ consisting of variables
$V=\{x_1, \ldots, x_n, y_1, \ldots, y_m\}$ with a proper partial
assignment $t$ over the set $\{y_1, y_2, \ldots, y_m\}$ is given.
The question is:

``Is there a partial assignment $t'$ (not necessarily proper) over
the set $\{x_1, \ldots, x_n\}$ such that $\phi$ has a unique
proper assignment $r$ which satisfies $r(x_i)=t'(x_i)$ for every
$1 \le i \le n$?''

Note that since $t$ is a proper partial assignment, if such $t'$
exists, then $r(y_j)=t(y_j)$ for $1 \le j \le m$.
\begin{itemize}
\item $\exists \Unique_* \ k\SAT$ \item[]{\sc Instance:} A $k$CNF,
$\phi(\bx,\by)$ and a proper partial assignment $t$ over the set
of the variables $y_j$.
\\
{\sc Question:} Is $\exists \bx \Unique_t \by \ \phi(\bx,\by)$?
\end{itemize}
\begin{theorem}
\label{unique4} The $\exists \Unique_* \ 4\SAT$ problem is
\sg-complete.
\end{theorem}
\begin{proof}
The problem is in \sg. To prove the completeness, we give a
reduction from  $\exists {\not\exists} \ 3\SAT$. Consider a 3CNF,
 $\phi(\bx,\by)$ and the problem $\exists \bx {\not\exists} \by \ \phi(\bx,\by)$.
We construct an instance of $\exists \Unique_* \ 4\SAT$, a $4$CNF
$\mu$ with a proper partial assignment $t$, as in the following.
The expression $\mu$ has all of the variables of $\phi$ plus one
more variable $z$. Let $C_1, C_2, \ldots, C_n$ be the clauses of
$\phi(\bx,\by)$. Then

\begin{equation}
\label{mu}
 \mu(\bx,\by,z)=(C_1 \vee z) \wedge (C_2 \vee z) \wedge \ldots
\wedge (C_n \vee z) \wedge (\bar z \vee y_1) \wedge (\bar z \vee
y_2) \wedge \ldots (\bar z \vee y_m)
\end{equation}

The partial assignment $t(z)=\true$ and $t(y_j)=\true$ ($1\le j\le
m$) is given. This partial assignment is proper because there
exists a true literal in every clause among $z$ and $y_j$'s. For
every proper assignment $u$ of $\mu(\bx,\by,z)$, if $u(z)=\true$,
then $u(y_j)=\true$ ($1 \le j \le m$). If $u(z)=\false$, then by
ignoring the variable $z$ in $u$, $u$ is a proper assignment of
$\phi(\bx,\by)$, and vice versa.
 So
$\exists \bx {\not\exists} \by \ \phi(\bx,\by)$ if and only if
 $\exists \bx \Unique_t (\by, z) \ \mu(\bx,\by,z)$.
\end{proof}

Next, we modify the proof of Theorem~\ref{unique4} so that we can
conclude that $\exists \Unique_* \ 3\SAT$ is \sg-complete.
Consider $\mu(\bx,\by,z)$, defined in~(\ref{mu}). In
$\mu(\bx,\by,z)$ every clause of the form $(\bar z \vee y_j)$ has
two literals. But a clause of the form $(C_i \vee z)$ has four
literals. Suppose $C_i=a_1 \vee a_2 \vee a_3$, where $a_1, a_2,
a_3$ are literals. We replace each clause $(C_i \vee z)$ in $\mu$
by $C'_i$ defined as follows,
$$
C'_i=(a_1 \vee a_2 \vee v_i) \wedge (a_3 \vee z \vee \bar v_i)
\wedge$$
$$(\bar a_1 \vee \bar z \vee v_i) \wedge (\bar a_2 \vee
\bar z \vee v_i) \wedge (\bar a_1 \vee \bar a_3 \vee v_i)\wedge
(\bar a_2 \vee \bar a_3 \vee v_i),$$
where $v_i$'s are new variables, and call the new expression as
$\mu'(\bx,\by,\bv,z)$. Thus
$$\mu'(\bx,\by,\bv,z)=C'_1 \wedge \ldots \wedge C'_n  \wedge (\bar z \vee y_1) \wedge (\bar z \vee
y_2) \wedge \ldots (\bar z \vee y_m).$$

 Define the partial
assignment as $u(z)=\true$, $u(y_i)=\true$, $u(v_i)=\true$ for
$1\le i\le m$.

The following three observations imply that $\exists \bx \Unique_t
(\by,z) \ \mu(\bx,\by,z)$ if and only if $\exists \bx \Unique_u
(\by,\bv,z) \ \mu'(\bx,\by,\bv,z)$.

\begin{itemize}
  \item[(a)] $u(z)=\true$, $u(y_j)=\true$ ($1\le j\le m$),
        and $u(v_i)=\true$ ($1\le i \le n$) is a proper partial assignment of $\mu'(\bx,\by,\bv,z)$.

  \item[(b)] Every truth assignment to $a_1$, $a_2$, $a_3$, and $z$
       which assigns a true value to at least one of them
       is extended uniquely to a proper assignment of $C'_i$.

  \item[(c)] Since every assignment which assigns   a false value to $a_1$, $a_2$, $a_3$, and
       $z$ simultaneously
        is not a proper assignment of $C'_i$, every proper assignment of $\mu'(\bx,\by,\bv,z)$ leads to a
        proper assignment of $\mu(\bx,\by,z)$ by ignoring the values of
        $v_i$'s.
\end{itemize}

Note that any proper subset of the clauses of $C'_i$ does not
satisfy these properties. For example consider the assignment
$t(a_1)=\false$, $t(a_2)=\true $, $t(a_3)=\true $, $t(z)=\false$.
In this case regardless of what value is assigned to $v_i$ the
first five clauses are satisfied, and the last clause is necessary
to fix the value of $v_i$.

We conclude the following theorem from (a), (b), and (c).
\begin{theorem}
\label{unique3} The $\exists \Unique_* \ 3\SAT$ problem is
\sg-complete.
\end{theorem}

In the proof of Theorem~\ref{unique3} the problem $\exists
{\not\exists} \ 3\SAT$ is reduced to $\exists \Unique_* \ 3\SAT$.
In that proof by assuming that there are no variables $x_i$'s in
$\phi(\bx,\by)$ (i.e. the number of variables after the first
quantifier of $\exists {\not\exists} \ 3\SAT$ is zero), we can
obtain a reduction form ${\not\exists} \ 3\SAT$ to the problem
which asks whether a given proper assignment of a 3CNF is its only
proper assignment. This problem is a restriction of Q1-$3\SAT$ in
which $D$, the set which is asked to be the defining set, is the
empty set. Since ${\not\exists} \ 3\SAT$ is \CoNP-complete, we
have:

\begin{theorem}
Q$1$-$3\SAT$ is \CoNP-complete.
\end{theorem}

The next theorem determines the computational complexity of
Q2-$3\SAT$.

\begin{theorem}
Q$2$-$3\SAT$ is \sg-complete.
\end{theorem}
\begin{proof}
The problem is in \sg. We reduce $\exists \Unique_* \ 3\SAT$ to
this problem. Let $\exists \bx \Unique_t \by \ \mu(\bx,\by)$ be an
instance of $\exists \Unique_* \ 3\SAT$, where $\mu(\bx,\by)$ is a
3CNF with variables $x_1, x_2, \ldots, x_k$ and $y_1, y_2, \ldots,
y_m$, and $t$ is a proper partial assignment over variables $y_j$.
We construct an instance of Q2-3\SAT, a $3$CNF $\phi$ with a
proper assignment $t'$, such that $({\cal P}(\phi),t')$ has a
defining set of size at most $k$, the number of the variables
$x_i$, if and only if $\exists \bx \Unique_t \by \ \mu(\bx,\by)$.
In the following we describe how $\phi$ is obtained from
$\mu(\bx,\by)$.

For every $1 \le i \leq k$, consider two new variables $v_i$ and
$v'_i$, and replace every $x_i$ in each clause of $\mu(\bx,\by)$
with $v_i$ and every $\bar x_i$ with $v'_i$.

For every $1 \le j \le m$, a literal $a_j$ is defined as follows.
If $t(y_j)=\false$, then $a_j$ is $y_j$ and otherwise $a_j$ is
$\bar y_j$. We add the following clauses to the expression in
which $w_i$ are new variables.

\begin{equation}
\label{w_i} (a_1 \vee a_2 \vee w_1) \wedge (\bar w_1 \vee a_3 \vee
w_2) \wedge \ldots \wedge (\bar w_{m-2} \vee a_m \vee w_{m-1})
\end{equation}

Note that by setting $y_j$'s according to the given assignment
$t$, $w_i$'s are forced to take the truth value $\true$. The
following clauses are also added to the expression.
\[(\bar w_{m-1} \vee v_1 \vee \bar v'_1) \wedge
  (\bar w_{m-1} \vee \bar v_1 \vee v'_1) \wedge \ldots \wedge
  (\bar w_{m-1} \vee v_k \vee \bar v'_k) \wedge
  (\bar w_{m-1} \vee \bar v_k \vee v'_k) \]
Call this new 3CNF, $\phi(\bv,\bv',\by,\bw)$. Let $t'$ be the
assignment
 $t'(v_i)=t'(v'_i)=\false$ ($1 \le i \le k$),
$t'(w_i)=\true$  ($1\le i \le m-1$), and $t'(y_j)=t(y_j)$  ($1\le
j \le m$).  Note that $t'$ is a proper assignment of $\phi$.

We claim that $({\cal P}(\phi),t')$ has a defining set of size at
most $k$, if and only if $\exists \bx \Unique_t \by \
\mu(\bx,\by)$.

Suppose that $\exists \bx \Unique_t \by \ \mu(\bx,\by)$. This
means that there is a partial assignment $u$ over $x_1, x_2,
\ldots, x_k$ such that the only proper values for $y_j$ are the
values that are assigned to them by the partial assignment $t$. If
$u(x_i)=\true$, we choose $(v'_i,\false)$, and if $u(x_i)=\false$,
we choose $(v_i,false)$.  Call this set $S$. We claim that $S$ is
a defining set of $({\cal P}(\phi),t')$.

Suppose that $S$ is not a defining set of  $({\cal P}(\phi),t')$.
Then there is a proper assignment $t'' \neq t'$ which is an
extension of $S$. Since $t''(v_i)=\false$ and $t''(v'_j)=\false$
for $v_i ,v'_j \in \supp(S)$, it can be easily seen that the
assignment $r$ defined as $r(x_i)=u(x_i)$ ($1 \le i \le k$) and
$r(y_j)=t''(y_j)$ ($1 \le j \le m$) is a proper assignment to the
variables of $\mu(\bx,\by)$ which is a contradiction. So all $y_j$
take the values that are assigned to them by  the assignment $t$.
Hence $w_i$'s are true for all $1 \leq i \leq m-1$. Since the two
clauses $(\bar w_{m-1} \vee v_i \vee \bar v'_i)$ and $(\bar
w_{m-1} \vee \bar v_i \vee v'_i)$ are in $\phi$, and exactly one
of $v_i$ or $v'_i$ is in $\supp(S)$, the value of the other one is
also determined to be false, and this is the assignment $t'$.

Next suppose that $({\cal P}(\phi),t')$ has a defining set $S$ of
size at most $k$. Then for every $1\leq i\leq k$, at least one of
$v_i$ or $v'_i$ is in $\supp(S)$. Otherwise we can change the
values of both $v_i$ and $v'_i$ to true, and still have a proper
assignment. So a defining set of size at most $k$ includes exactly
one of $(v_i,\false)$ or $(v'_i,\false)$ for every $1 \leq i \leq
k$. Let $u$ be a partial assignment of $\mu(\bx,\by)$ such that
$u(x_i)=\true$ if $v_i\in \supp(S)$, and $u(x_i)=\false$ if $v'_i
\in \supp(S)$.

We claim that $\mu(\bx,\by)$ has a unique proper assignment $r$
such that $r(x_i)=u(x_i)$ for every $1 \le i \le k$. Suppose that
there is a proper assignment $r$ for $\mu(\bx,\by)$ such that
$r(x_i)=u(x_i)$ for all $1\le i \le k$, but there exists at least
one $1 \le i_0 \le m$ such that $r(y_{i_0}) \neq t(y_{i_0})$.

Consider $\phi(\bv,\bv',\by,\bw)$, and let $r'(y_j)=r(y_j)$ ($1
\le j \le k$). Since $r(y_{i_0}) \neq t(y_{i_0})$, it is possible
to assign values $r'(w_i)$ ($1\le i \le m-1$) such that
$r'(w_{m-1})=\false$ and the clauses in~(\ref{w_i}) are true.

Note that \emph{exactly} one of $v_i$ or $v'_i$ is in $\supp(S)$.
For every $1 \le i \le k$, if $v_i \in \supp(S)$, then define
$r'(v_i)=\false$, $r'(v'_i)=\true$; and if $v'_i \in \supp(S)$,
then define $r'(v_i)=\true$, $r'(v'_i)=\false$.

Since $t(w_{m-1})=\false$, the values assigned by $r'$ do not make
$(\bar w_{m-1} \vee v_i \vee \bar v'_i) \wedge (\bar w_{m-1} \vee
\bar v_i \vee v'_i)$ false.

Now all clauses are satisfied. So there exists another proper
assignment containing the defining set, which is a contradiction.
\end{proof}
\begin{theorem}
Q$3$-$3\SAT$ is \sg-complete.
\end{theorem}
\begin{proof}
The problem is in \sg. We give a reduction from Q2-$3\SAT$.
Consider an instance of Q2-$3\SAT$, a 3CNF $\phi$ with a proper
assignment $t$ and an integer $k$. Let the variables of $\phi$ be
$x_1, x_2, \ldots, x_n$. We add $n(k+1)$ new variables $y_{ij}$
($1\le i\le n$ and $1\le j\le k+1$). For every $x_i$, if
$t(x_i)=\true$, then we add the following clauses:
\[(\bar x_i \vee y_{i1}) \wedge (\bar x_i \vee y_{i2})\wedge \ldots
\wedge (\bar x_i \vee y_{i(k+1)}),\] and if $t(x_i)=\false$, then
we add the following clauses:
\[(x_i \vee y_{i1}) \wedge (x_i \vee y_{i2})\wedge \ldots
\wedge (x_i \vee y_{i(k+1)}).\]

The new 3CNF  consists of $\phi$ and these $n(k+1)$ new clauses.
Denote this 3CNF by $\phi'$. We claim that ${\cal P}(\phi')$ has a
defining set of size at most $k$, if and only if $({\cal
P}(\phi),t)$ has a defining set of size at most $k$. Every
defining set of $({\cal P}(\phi),t)$ is also a defining set of
${\cal P}(\phi')$, because the assignment $t$ forces all of the
$y_{ij}$ to take a true value.

Next suppose that there is a defining set of ${\cal P}(\phi')$
which fixes a proper assignment $t'$. For every $1 \le x \le n$,
if $t'(x_i) \neq t(x_i)$, then since it is possible to assign
every arbitrary values to $y_{i1}, y_{i2}, \ldots, y_{i(k+1)}$,
all these $k+1$ variables are in the defining set. Hence in every
defining set of size at most $k$, all $x_i$ take the same values
in $t'$ and $t$. Now, since $t'(x_i)=t(x_i)$, by fixing the value
of $x_i$, the values of $y_{ij}$'s are determined to be true, for
$1 \le j \le k+1$. So if $(y_{ij},t'(y_{ij}))$ is in the defining
set, then it is possible to replace it by $(x_i,t'(x_i))$. Thus a
defining set of size at most $k$ of ${\cal P}(\phi')$ can be
modified so that all its elements are in $\{(x_i,t'(x_i)):
i=1,\ldots,n\}$, and $t'(x_i)=t(x_i)$. This is also a defining set
of $({\cal P}(\phi),t)$.
\end{proof}

\section{Vertex Coloring}
For every graph $G$ with vertex set $V=\{ v_1, \ldots, v_n \}$,
every vertex coloring $c$ of $G$ is a function which maps every
vertex $v_i$ to a color $c(v_i)$. For every partial coloring $c$
of $G$, define $\supp(c)$ as the set of the vertices that $c$
assigns a color to them.
 Denote the family of all
$\chi(G)$-vertex colorings of $G$ by ${\cal C}(G)$.
In~\cite{Dailey} it is shown that the uniqueness of colorability
is \CoNP-complete. This implies  the following theorem.

\begin{theorem}
The problem Q$1$-\vc is \CoNP-complete.
\end{theorem}

In this section we show that both of the problems Q2 and Q3 for
this family are \sg-complete.
\begin{itemize}
\item Q2-\vc \item[] {\sc Instance:} A graph $G$ with a
$\chi(G)$-vertex coloring $c$, and an integer $k$.
\\
{\sc Question:} Does $({\cal C}(G),c)$ have a defining set of size
at most $k$?
\end{itemize}
\begin{itemize}
\item Q3-\vc \item[] {\sc Instance:} A graph $G$, and an integer
$k$.
\\
{\sc Question:} Does ${\cal C}(G)$ have a defining set of size at
most $k$?
\end{itemize}

\begin{theorem}
Q$2$-\vc is \sg-complete for graphs with $\chi=3$.
\end{theorem}
\begin{proof}
The problem is in \sg. To prove the completeness, we introduce a
reduction from Q$2$-$3\SAT$. Consider an instance of Q2-$3\SAT$: A
proper assignment $t$ of $\phi(x_1,x_2, \ldots, x_n)$ and an
integer $k$. We construct a graph $G_\phi$ with chromatic number
$3$ and a $3$-vertex coloring $c_t$ of $G_\phi$ such that $({\cal
P}(\phi),t)$ has a defining set of size at most $k$ if and only if
$({\cal C}(G_\phi),c_t)$ has a defining set of size at most $k+4$.

 We begin by considering a cycle of size $3$ with vertices $w_{0}$, $w_{1}$,
and $w_{2}$ which are connected to four vertices $w'_{1}, w'_{2},
w'_{3},$ and $w'_{4}$ as it is shown in Figure~\ref{Q2}(a). For
every variable $x_i$, add two vertices $u_{x_i}$ and $u_{\bar
x_i}$ and edges $\{ u_{x_i},u_{\bar x_i} \}$, $\{ u_{x_i},w_2\}$,
and $\{ u_{\bar x_i},w_2\}$ to the graph. This is illustrated in
Figure~\ref{Q2}(a).

\begin{figure}[ht]
\input{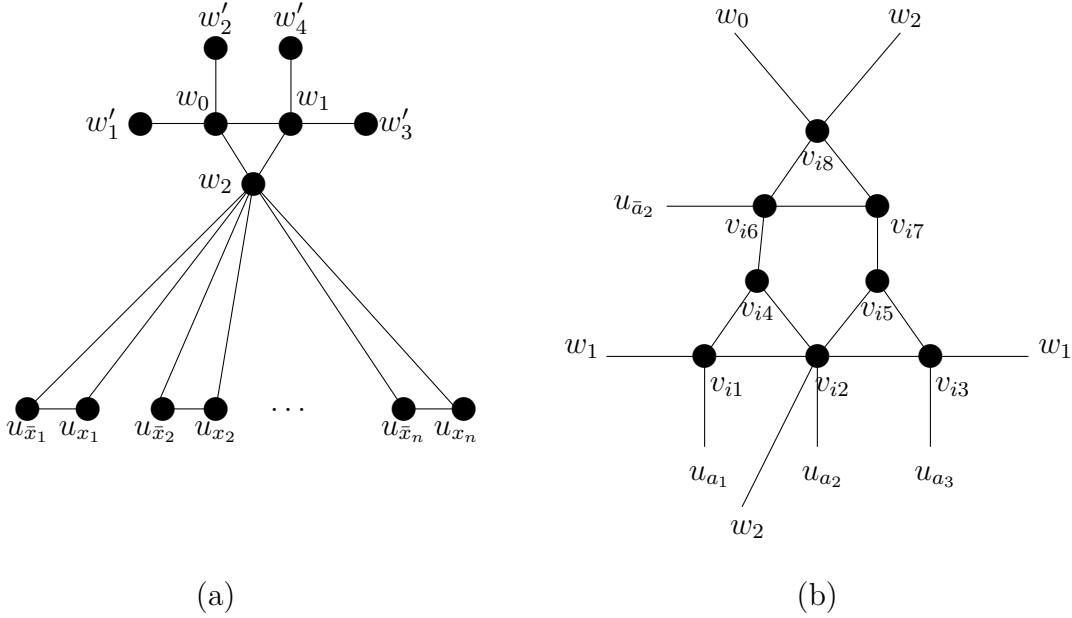}
\caption{\label{Q2}(a) The vertices $u_{x_i}$ and $u_{\bar x_i}$
are connected to $w_2$. (b) For every clause we add a copy of this
graph to $G_\phi$.}
\end{figure}

Consider a clause $C_i=(a_1 \vee a_2 \vee a_3)$ of $\phi$, where
$a_j$ ($j=1,2,3$) is a literal. Since $t$ is a proper assignment
of $\phi$, without loss of generality we can assume that
$t(a_2)=\true$. For every such clause, we add a copy of the graph
shown in Figure~\ref{Q2}(b) to the graph, and connect its vertices
to the other vertices as it is shown in Figure~\ref{Q2}(b). Notice
that $u_{a_j}$ ($j=1,2,3$) is one of the vertices $u_{x_1},
u_{x_2}, \ldots, u_{x_n}$ or $u_{\bar x_1}, u_{\bar x_2}, \ldots,
u_{\bar x_n}$. Call this new graph as $G_\phi$.

One can easily check that assigning a $3$-coloring $c_t$ to
$u_{a_1}$, $u_{a_2}$, and $u_{a_3}$ such that $c_t(w_{0})=0$,
$c_t(w_{1})=1$, and $c_t(w_{2})=2$ and also $c_t(u_{a_2})=1$
determines the colors of $v_{i1}, v_{i2}, \ldots, v_{i8}$
uniquely. Let $c_t$ be a $3$-coloring of $G_\phi$ defined as in
the following:

\begin{itemize}
\item $c_t(w_{0})=0$, $c_t(w_{1})=1$, and $c_t(w_{2})=2$.

\item $c_t(w'_{1})=1, c_t(w'_{2})=2, c_t(w'_{3})=0$, and
$c_t(w'_{4})=2$.

\item For every $1 \le i \le n$ if $t(x_i)=\true$, then
$c_t(u_{x_i})=1$ and $c_t(u_{\bar x_i})=0$, and otherwise
$c_t(u_{x_i})=0$ and $c_t(u_{\bar x_i})=1$.

\item Colors of $v_{ij}$ are determined uniquely by the colors of
the vertices above.
\end{itemize}

The vertices $w'_{1}, w'_{2}, w'_{3}$, and $w'_{4}$ are in every
defining set (otherwise we can change their colors). The colors of
these four vertices determine the colors of $w_0$, $w_1$, and
$w_2$ uniquely.

We claim that the size of the smallest defining set of $({\cal C}
(G_\phi),c_t)$ is equal to the size of the smallest defining set
of $({\cal P}(\phi),t)$ plus $4$.  Note that any partial coloring
which only assigns $0$ or $1$ to $u_{a_1},u_{a_2},u_{a_3}$ and
does not assign $0$ to all of them can be extended to a proper
coloring of the graph in Figure~\ref{Q2}(b). Moreover if all the
vertices $u_{a_1}, u_{a_2}, u_{a_3}$ are colored by $0$, then it
can be easily seen that $v_{i8}$ is also forced to be colored by
$0$. Since $v_{i8}$ is connected to $w_{0}$ and $w_{2}$, $G_\phi$
admits a $3$-coloring, if and only if $\phi$ has a proper
assignment.

Suppose $({\cal C}(\phi),t)$ has a defining set consists of $k$
variables $x_{i_1}, x_{i_2}, \ldots, x_{i_k}$. Then assigning
colors of $k+4$ vertices $w'_{1}, w'_{2}, w'_{3}, w'_{4}$ and
$u_{x_{i_1}}, u_{x_{i_2}}, \ldots ,u_{x_{i_k}}$ constitutes a
defining set of $({\cal C}(G_\phi),c_t)$.

Next suppose that $S$ is the smallest defining set of $({\cal
C}(G_\phi),c_t)$. Then $w'_{1}$, $w'_{2}$, $w'_{3}$, $w'_{4}$ are
in $\supp(S)$. By assigning the colors of these vertices the
colors of $w_{1}$, $w_{2}$, $w_{3}$, and all $v_{i8}$'s are
determined uniquely. It can be verified easily that for every
clause $C_i = (a_1, a_2, a_3)$ of $\phi$, since $c_t(u_{a_2})=1$,
the colors of $v_{i1}, v_{i2}, \ldots, v_{i7}$ are determined
uniquely by fixing the color of $u_{a_2}$, and the color of
$u_{x_i}$ determines the color of $u_{\bar x_i}$. Hence we can
assume that $\supp(S)$ contains $w'_1$, $w'_2$, $w'_3$, $w'_4$,
and some of $u_{x_i}$. Using the fact that any partial coloring
which only assigns $0$ or $1$ to $u_{a_1},u_{a_2},u_{a_3}$ and
does not assign $0$ to all of them can be extended to a proper
coloring of the graph in Figure~\ref{Q2}(b), we conclude that the
corresponding variables of these $u_{x_i}$ constitute a defining
set of $({\cal C}(\phi),t)$.
\end{proof}

\begin{theorem}
Q$3$-\vc is \sg-complete for graphs with $\chi=3$.
\end{theorem}
\begin{proof}
The problem is in \sg. We give a reduction from Q2-\vc when
$\chi=3$. Consider an instance $({\cal C}(G),c)$ of Q2-\vc, where
$G$ is a graph and $c$ is a $3$-vertex coloring of $G$. Assume
that the range of $c$ is the set $\{0,1,2\}$. An integer $k$ is
given, and it is asked that ``Is there a defining set of size at
most $k$ for $({\cal C}(G),c)$?'' We construct a new graph $H$ as
follows:
\begin{enumerate}
   \item First let $H$ be the disjoint union of $G$ and a cycle $w_0w_1w_2$ of size 3. Then
   \item for every vertex $u_i$ of $G$, let $c_1$ and $c_2$ be the
         two colors other than $c(u_i)$. Add $2k+2$ vertices
         $v_{u_i, c_j, 1}, v_{u_i, c_j, 2}, \ldots, v_{u_i, c_j, k+1}$
         ($1\le j \le 2$) to $H$. For every $1\le t \le k+1$,
         connect $v_{u_i, c_j, t}$ to both $u_i$ and $w_{c_j}$.
         (Notice that $w_{c_j}$ is one of $w_0$, $w_1$, or $w_2$.)
   \item Add four new vertices $w'_1$, $w'_2$, $w'_3$, and $w'_4$ to
         $H$, and connect $w'_1$ and $w'_2$ to $w_0$, and also $w'_3$
         and $w'_4$ to $w_1$.
\end{enumerate}

Now we claim that ${\cal C}(H)$ has a defining set of size at most
$k+4$ if and only if $({\cal C}(G),c)$ has a defining set of size
at most $k$.

First consider a defining set of size at most $k$ for $({\cal C}
(G),c)$, say $D$. If we fix the colors of the vertices in $D$ and
assign the colors 1 to $w'_1$, 2 to $w'_2$, 0 to $w'_3$, and 2 to
$w'_4$, then these $k+4$ vertices constitute a defining set of
${\cal C}(H)$.

 Next suppose that $D$ is the smallest defining set of ${\cal C}(H)$
which has at most $k+4$ vertices. Without loss of generality
assume that in the extension of $D$ to a $3$-vertex coloring $c'$
of $H$, $w_i$ ($0 \le i \le 2$) is colored by $i$. Since the
degrees of $w'_1$, $w'_2$, $w'_3$, and $w'_4$ are equal to one,
they are in $\supp(D)$. Suppose that in the extension of $D$ to a
$3$-coloring of $H$, a vertex $u_i$ of $G$ is colored by $c'(u_i)$
which is not equal to $c(u_i)$. The vertices $v_{u_i, c'(u_i), t}$
($1 \le t \le k+1$) are only connected to $u_i$ and $w_{c'(u_i)}$.
Since these two vertices are colored by the same colors, all these
$k+1$ vertices are in the defining set, and with the four vertices
$w'_i$, the size of the defining set is at least $k+5$. Since $D$
is of size at most $k+4$, for every vertex $u_i$,
$c'(u_i)=c(u_i)$.

We can suppose that $w'_1$ and $w'_2$ (and so $w'_3$ and $w'_4$)
are colored by different colors. Otherwise by changing the color
of $w'_2$, (and so $w'_4$),  $D$ still remains a defining set of
${\cal C}(H)$. Since $w'_1$ and $w'_2$ and also $w'_3$ and $w'_4$
are colored by different colors, they determine the colors of
$w_1$, $w_2$, and $w_3$ uniquely. Therefore since $D$ is the
smallest defining set, none of $w_1$, $w_2$, and $w_3$ is in
$\supp(D)$. Also if a vertex $v_{u_i, c_j, t}$ ($c_j \neq c(u_i)$)
is in $D$, then we can replace it with $u_i$. Now, if we remove
the four vertices $w'_i$ from the defining set, we obtain a
defining set of $({\cal C} (G),c)$.
\end{proof}
\section*{Acknowledgements}
The authors wish to thank A. Daneshgar for many useful comments
and suggestions in preparing this paper, and E.S. Mahmoodian and
P. Afshani for many valuable discussions towards the results of
this paper. We would also like to thank Oleg Verbitsky for
pointing out some errors in the draft version of this paper.
\bibliographystyle{plain}
\bibliography{sr36}
\end{document}